\documentclass{aastex631}

\begin{document}

\title{Occultation of microlensed images by circum-lens disks: Implications for surveys}

\author[0000-0002-0786-7307]{Elizabeth  Bailey}
\affiliation{UT San Antonio Department of Earth and Planetary Sciences}
\affiliation{UT San Antonio Department of Physics and Astronomy \\
One UTSA Circle \\
San Antonio, TX 78249, USA}



\begin{abstract}
Microlensing surveys typically apply automated criteria to exclude lightcurves that depart from time-symmetric and achromatic Paczy\'{n}ski curves. In surveys using such automated detection pipelines, occultation of microlensed images by circum-lens material could lead to underdetection or misclassification of ring- or disk-bearing compact object populations. This work characterizes the family of lightcurve morphologies by which uniform thin circum-lens disks of varying size and orientation, and low mass relative to their host lens, can intersect the ray paths produced during microlensing events. These morphology classes are paired with squared residuals relative to an unocculted lightcurve. We find that substantial deviations from an unocculted microlensing curve can occur across a wide range of parameter space. This effect holds relevance for free-floating planets and brown dwarfs where disks or ring systems can be present. Another potential consideration is that disk-shaped accretion flows onto primordial black holes (PBHs) during the cosmic dark ages could have resulted in remnant quiescent disks of baryonic material. Pipelines should be designed to account for disk-bearing lens populations, opening previously underutilized avenues to potentially observe disks and rings in these contexts and ensuring populations are accurately characterized. 
\end{abstract}

\section{Introduction} \label{sec:intro}

During a gravitational lensing event \citep{Einstein1936}, magnification of the source from the standpoint of the observer produces a time-symmetric and achromatic Paczy\'{n}ski lightcurve \citep{Paczynski}, which is selected for in automated microlensing surveys \citep{Alcocketal2000, Bond2001,Tisserand2007, Udalski2015}. Scenarios that cause microlensing lightcurves to deviate from the idealized Paczy\'{n}ski curve have been widely considered \citep{Paczynski1996}, including compound lenses or lenses orbited by planets \citep{MaoPaczynski1991, GouldLoeb, GriestSafizadeh1998, Hanetal2001, Rattenburyetal2003} and effects caused by acceleration or parallax \citep{Gould1992, Smithetal2003}. A related consideration is the use of gravitational lensing signals to infer features of the source---for example, the properties of accretion disks surrounding source quasars \citep{GouldEscude1997, Blackburne, Jiminez, HagenFian2026} or the features of source stars and their associated planetary systems \citep{GraffGaudi2000, LewisIbata2000, Gaudietal2003}. Due to the broad variety of possible deviations from an idealized microlensing lightcurve, selection bias is of general concern when automated detection pipelines are applied \citep[e.g.,][]{Kochanek2004}. 

Occultation of microlensing signals has been a related consideration, whether by clouds \citep{Nieuwenhuizen,Bozzaetal2002} or a spherical lens having adequate physical diameter to occult the lensed images \citep{Bromley1996, Algol2002, Takahashi2003}. This work draws attention to the case where circum-lens material, such as a disk or ring system, exists at orbital positions intersecting ray paths of the gravitationally lensed images, causing substantial suppression or distortion of lightcurves. Note that this mechanism involves physical intersection of the disk with the lensed ray paths, and could occur for disks having low mass compared to the lens. This class of mechanism has previously been considered for the case of envelopes surrounding strongly interacting binary lenses \citep{BozzaMancini2002} and as a means to detect circumstellar disks around F, G, K stars acting as lenses---an approach viewed as having limited observational utility in that context \citep{Hundertmark}. However, whether disks or rings around single compact lenses could cause microlensing events to be missed by survey pipelines does not appear to be widely considered at this time. Here we examine this possibility, which is of time-sensitive importance given the imminent Nancy Grace Roman Space Telescope Galactic Bulge Time-Domain Survey (GBTDS), designed to use high-cadence microlensing observations to characterize compact object populations \citep{Johnsonetal2020,DeRocco2024}.

Astrophysical disks commonly result from angular momentum conservation during gravitational collapse. In the context considered here, microlensing could help to address properties of disks in settings that otherwise challenge observations. This work provides a systematic exploration of the space of occultation geometries and respective lightcurves producible through this mechanism, offering a basis for application to microlensing surveys. We then discuss some potentially relevant astrophysical contexts. First, we will consider disk- or ring-bearing free-floating planets and brown dwarfs. This will be followed by a brief and conditional consideration of the potential for primordial black holes (PBHs) to host long-term stable disks and the optical properties of molecular hydrogen ice as they relate to this setting.

\section{Disk occultation geometry and parameter space of model lightcurves}\label{sec:geometries}

\subsection{Disk occultation model}
For a point lens, the lightcurve is the sum of contributions from two lensed images of the background source that sweep along separate trajectories in the viewing plane, with one interior and the other exterior to the Einstein ring \citep{Paczynski}. In the case of microlensing, these images are not spatially resolved. However, their separate positions become relevant to observations if a disk surrounding the lens is physically present at a position that intersects the ray path of either image at specific points along the image's trajectory. This mechanism requires the disk to either attenuate the light or divert it, but the mass of the disk can be negligible compared to the lens and produce the effect considered here because the mechanism is physical intersection of the light path rather than gravitational influence of the disk.

We model the limiting case of a thin circular disk having negligible mass compared to the lens, and treated as extending all the way to the lens. For simplicity in illustrating the space of lightcurve geometries, intersection of the disk with the ray path of either image is treated as causing complete extinction of that image for the duration of occultation. Because the gravitational deflection angles due to typical microlensing events are extremely small, the light paths are treated as locally straight lines through the lens plane. Occultation is approximated as occurring where each image path intersects the projection of the circum-lens disk onto the lens plane. The source is approximated as a point source. 

Microlensing lightcurve morphologies were classified by a doublet [$n_{\text{interior}}$, $n_{\text{both}}$], where $n_{\text{interior}}$ is the number of distinct time segments during which only the interior lensed image is occulted, and $n_{\text{both}}$ is the number of time segments during which both lensed images are simultaneously occulted. It is never the case that only the exterior image is occulted at a given time by a lens-encircling and radially symmetric disk, because simultaneous interior and exterior images occur on a straight line through the origin. Lightcurve morphologies were classified for: disk radius $A$ ranging from $0$ to $3r_E$, where $r_E$ is the Einstein radius; disk orientation $\theta$ defined as the angular offset between the long axis of the projected ellipse and the axis of minimum separation of the source and lens (degenerate for a face-on disk); and disk tilt varying relative to the lens plane.

For each occulted lightcurve, corresponding total squared residuals were calculated as $\int_{-10} ^{10}(m_{\text{Pac}} - m_{\text{occ}})^2 dx$, where $m_{\text{Pac}}$ is the ideal unocculted Paczy\'{n}ski magnification curve, $m_{\text{occ}}$ is the occulted magnification curve, and $x$ is the horizontal position of the source in units of $r_E$ at each modeled time step. This metric was intended as a proxy for each modeled lightcurve's deviation from an unocculted lightcurve, analogous to a chi-squared metric but without assumptions about noise or cadence.

\subsection{Phase space of occulted lightcurve morphologies}

\begin{figure*}
\centering
\includegraphics[width=0.8\columnwidth]{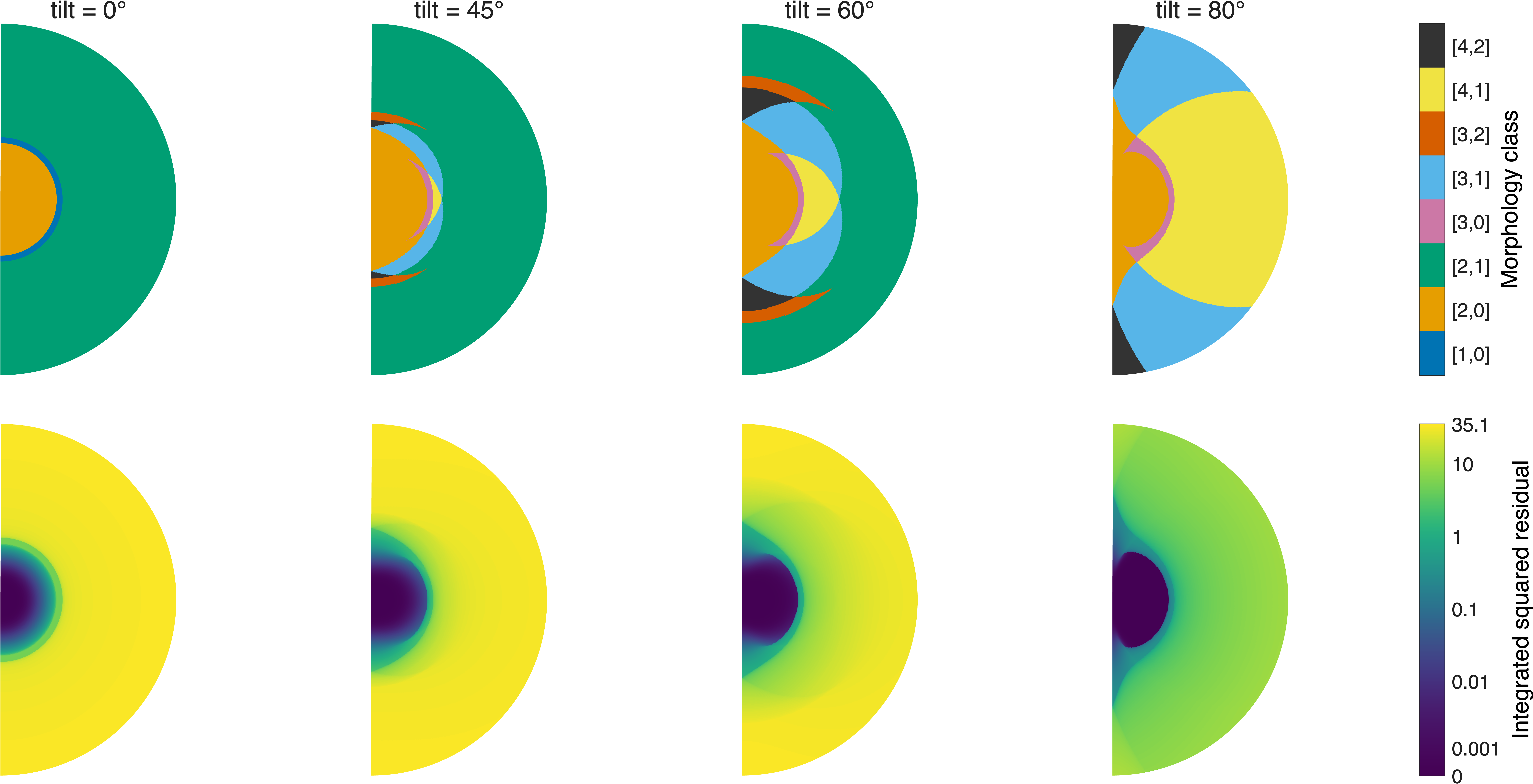}
\caption{Paired morphology space and squared residuals for disk-occulted lightcurves where the impact parameter is $0.1r_E$. Each polar plot shows disk radius $A = 0$ to $3r_E$, and angle $\theta$ denotes the angular offset of the disk's projected major axis from the axis of minimum separation between the source and lens (illustrated in Figure \ref{fig3}). Each pair of plots corresponds to a different disk tilt relative to the lens plane: $0$ (face-on), $45$, $60$, and $80$ (nearly edge-on) degrees. The top row maps occultation morphology classes defined by the number of time segments in which either only the interior image is blocked or both images are blocked. The bottom row shows corresponding total squared residuals.}
\label{fig1}
\end{figure*}

\begin{figure*}
\centering
\includegraphics[width=0.8\columnwidth]{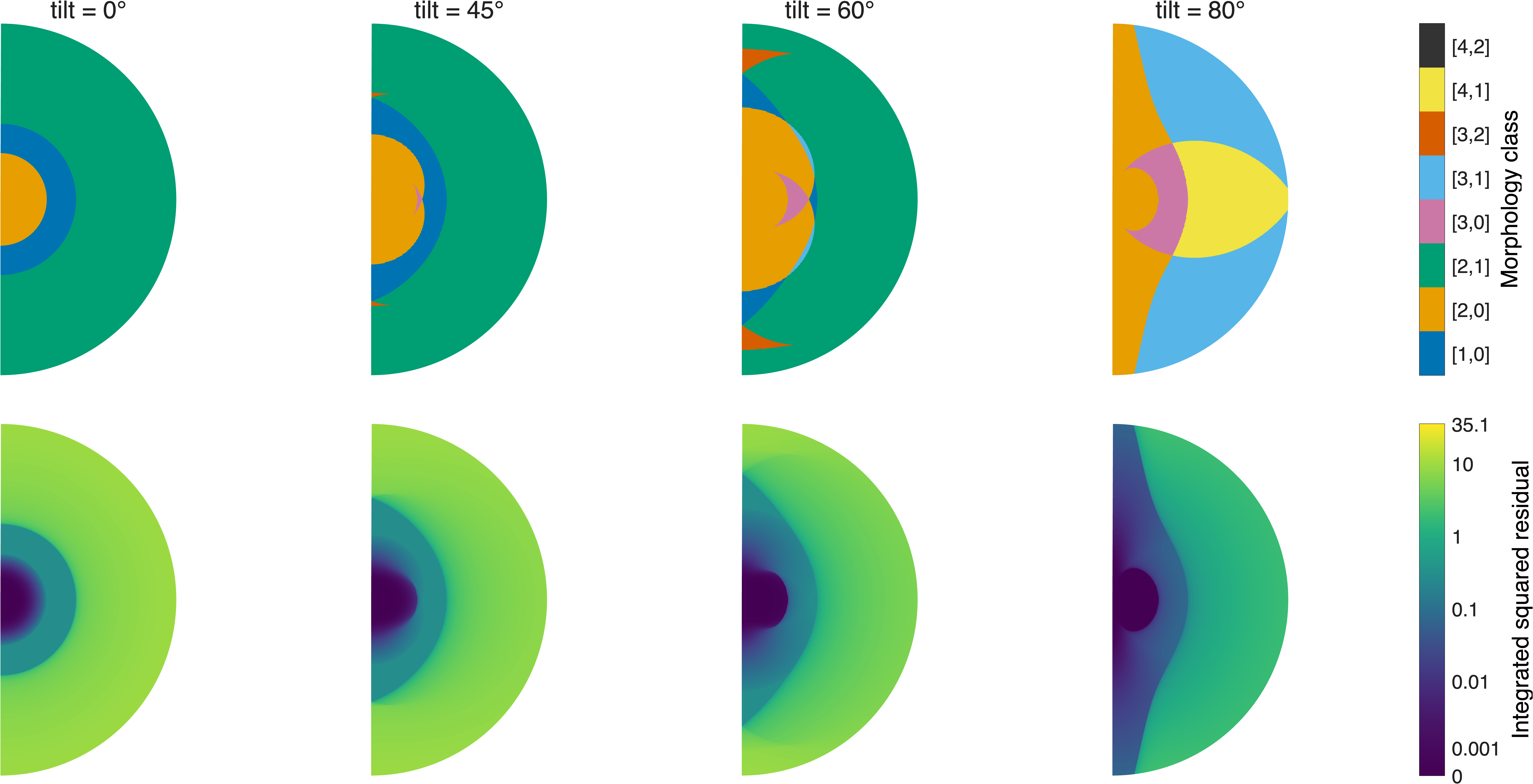}
\caption{Paired morphology space and squared residuals for disk-occulted lightcurves where the impact parameter is $0.5r_E$, analogous to the $0.1r_E$ impact parameter case given in Figure \ref{fig1}.}
\label{fig2}
\end{figure*}

For a microlensing impact parameter of $0.1r_E$ and disk tilts relative to the lens plane of $0^{\circ}$ (face-on), $45^{\circ}$, $60^{\circ}$, and $80^{\circ}$ (nearly edge-on),  Figure \ref{fig1} shows the parameter space of disk occultation morphologies, paired with the total squared residuals mapped across the same parameter space of disk radius relative to $r_E$ and orientation. Figure \ref{fig2} shows the analogous parameter space for an impact parameter of $0.5r_E$. Eight lightcurve morphology classes were identified. These classes are illustrated in Figure \ref{fig3} for the example case of impact parameter $0.1r_E$, alongside an illustration of corresponding disk projections onto the lens plane and paths of the two microlensed images.

\section{Discussion}
\begin{figure*}
\includegraphics[width=\columnwidth]{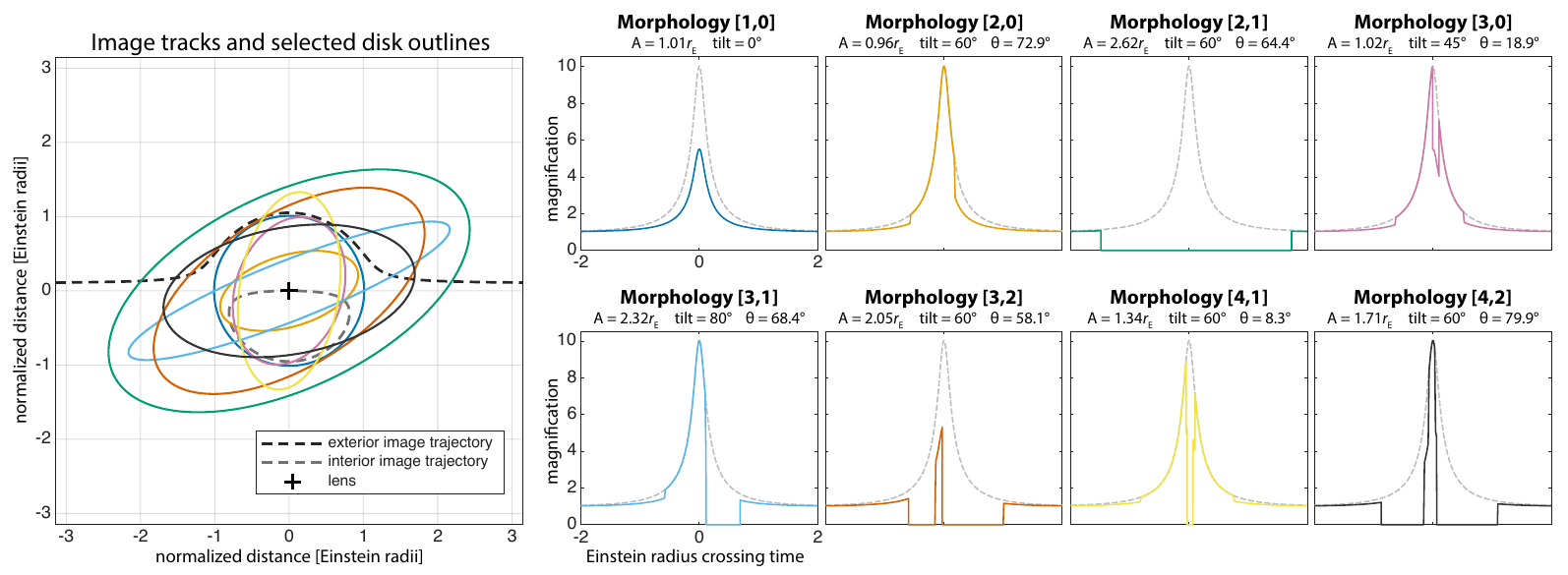}
\caption{Representative microlensing lightcurves produced by disk occultation. Colors correspond to the morphology space illustrated in Figures \ref{fig1} and \ref{fig2}. \textit{Left:} Outlines of tilted disks projected in the lens plane. Black and grey dashed lines represent the two lensed image paths. \textit{Right:} Eight corresponding occulted lightcurves. Unocculted Paczy\'{n}ski curves are shown with grey dashed lines.}
\label{fig3}
\end{figure*}

The eight occultation morphologies identified in this work have squared residuals ranging from relatively small, in the case of a physically small disk blocking only the interior image during the start and end of the microlensing event (designated as [2,0]; Figures \ref{fig1}--\ref{fig3}), to relatively large, in the case of large or face-on disks. When the disk is tilted relative to the lens plane, more complex geometry emerges in the morphology space due to loss of symmetry. Many of the results deviate profoundly from an unocculted lightcurve. A single disk can create repeated occultations leading to lightcurves with asymmetry and complex structure. Some details of each morphology class, as illustrated in Figure \ref{fig3}, are now discussed.

Full occlusion of both images across essentially the entire event is designated as morphology [2,1] because there is one time interval during which both images are occulted, which is flanked at the beginning and end by two intervals where only the interior image is occulted. A disk that is small relative to $r_E$ produces morphology [2,0], where the interior image is occulted at the beginning and end of the event. This scenario produces a notch-like dip on the flanks of the lightcurve peak (Figure \ref{fig3}).

An especially interesting occultation morphology is the class designated [1,0]. In this morphology class, the disk blocks the entire path of the interior image, but never the exterior image, resulting in a symmetric curve with a lower observed magnification than the expected unocculted curve. For a smaller impact parameter such as the $0.1r_E$ case illustrated here, this configuration requires a near-perfect size and alignment of a nearly face-on disk so that it occults only the interior image in its entirety. However, it becomes more likely to occur for larger impact parameters as the corresponding area of the parameter space increases with impact parameter due to increasing separation of the two lensed images from the Einstein ring (Figures \ref{fig1}, \ref{fig2}). Given their symmetry and potential achromaticity, such curves could be misclassified as non-occulted microlensing events, potentially leading to estimates of erroneously wide impact parameters relative to $r_E$, or short $r_E$ crossing times. Either interpretation could map to an erroneously low lens mass. Another potential concern with this morphology class is that the resulting lower magnification might not exceed the threshold of detectability for a given survey, especially in the case of larger microlensing impact parameters for which it occupies appreciable parameter space, and for which the magnification is smaller.

The relatively simple geometric cases of [2,1], [2,0], and [1,0] discussed above bear resemblance to geometries previously considered for the case of a large, opaque spherical lens object \citep{Takahashi2003}, with a major difference being that the disk orientation can break spherical symmetry and produce additional, asymmetric lightcurve morphologies in the cases [2,1] and [2,0] (Figure \ref{fig3}). Another way that this result differs from the previous geometries studied for an occulting spherical lens \citep{Algol2002, Takahashi2003} is that the lens objects of interest to Roman GBTDS are not generally expected to have physical radii comparable to or larger than the Einstein radius, whereas circum-lens disks or rings may plausibly exist at the relevant scales, as discussed in Sections \ref{planets} and \ref{holes}.

When the disk is tilted relative to the lens plane, additional geometric cases emerge compared to the case of a face-on disk (Figures \ref{fig1}, \ref{fig2}). For example, morphology [3,0] results in an asymmetric lightcurve with a notch. This case produces a lightcurve having a qualitatively similar appearance to what would be expected for a planetary caustic (Figure \ref{fig3}). Accordingly, surveys should be prudent to distinguish between the two cases. On a related note, the case illustrated here provides an illustration of the general tendency for many of these lightcurves to jump between the unocculted signal from the additive contribution of both images, the signal from the exterior image plus the occulted interior image, and the case where both images are occulted. Surveys could utilize this expected behavior in identifying disk occultation events and distinguishing them from other effects such as caustics.

Similar to the cases discussed above, morphologies [3,1], [3,2], [4,1], and [4,2] can likewise appear severely disrupted. The examples shown in Figure \ref{fig3} reflect the maximally simple and extreme case where occultation causes full extinction of a background point source. Disks with annular rings or varying extinction could impart an even more complex morphology. This work offers a conceptual starting point to guide fitting any such observed lightcurves to inferred disk or ring structures. Of particular note is that, in addition to affecting magnification, the astrometric position would be affected by a disk occultation. Astrometry could therefore prove to be a very useful tool for identifying the lightcurves produced by this class of mechanism, distinguishing them from other effects, and fitting them to specific disk structures, especially for larger lenses for which astrometry is most feasible.

We next discuss a few potentially relevant astronomical contexts, ranging from relatively mundane expected cases of planetary-mass objects and brown dwarfs, to the undeniably speculative but potentially relevant case of primordial black holes.

\subsection{Free-floating planetary-mass objects and brown dwarfs}\label{planets}
Observations of free-floating planetary- or brown-dwarf-mass objects frequently yield spectroscopic features consistent with the presence of disks \citep{SeoScholz2025, Damianetal2025}, including objects as small as $\sim 5 M_{\text{Jup}}$ \citep{Langeveld}. The lifespans of these disks are not well-constrained. \cite{SeoScholz2025} showed that emission from the inner region of disks around free-floating planetary-mass objects can persist as long as $10$ Myr. However, once cooling progresses, such emission may lose its utility as a means of observation, leaving occultation and/or microlensing as potentially valuable observational techniques.

For a lens of mass $M$, the physical Einstein radius is $r_{E} = \sqrt{ \frac{4GM}{c^2} \frac{D_L (D_S-D_L)}{D_S} }$ \citep{Paczynski}. For a source in the Galactic bulge, $r_E$ is typically of order $0.1$ au for a Jupiter-mass lens and $0.4 - 1$ au across the brown dwarf mass range; for a Large Magellanic Cloud source, the Einstein radius is larger by a factor of a few. Estimates for the outer radius of ring- or disk-like structures around brown dwarfs range from under $10^{-3}$ au for a debris belt that was modeled to fit a spectral energy distribution of a young brown dwarf \citep{Zakhozhay}, to under $25$ au based on ALMA observations that did not spatially resolve the disk \citep{Testi}, to $70$ au or larger for several brown dwarf disks from spatially resolved ALMA observations \citep{Ricci}. Disks of this approximate scale would typically be capable of occulting microlensing signals.

Another consideration is Mamajek-class disks, the nature of whose central objects remains of uncertain classification. In 2007, the young star J1407 underwent a complex dimming for over $50$ days, initially interpreted as occultation by the structured ring system of a gravitationally bound companion \citep{Mamajeketal2012, VanWerkhoven, KenworthyMamajek2015}, which was deemed likely planetary- or brown-dwarf-mass based on a lack of detected emission in follow-up observations \citep{Kenworthyetal2015}. Several lines of evidence have since been interpreted to support the possibility that the ring-bearing occulter may have been an unbound object. Such occulting rings, if they belonged to a bound companion, would be challenged by stability limits \citep{Kenworthyetal2015, RiederKenworthy2016}. No recurring eclipse was found in archival or subsequent monitoring \citep{Menteletal2018}, and separate observations resulted in non-detection of any source consistent with a bound companion \citep{Kenworthyetal2020, Klaassen}. Similar dimmings of other stars have in some cases been interpreted as potential eclipses by similar hypothesized ringed objects \citep{Pramono, Fores-Toribio, Zakamska, Osbornetal2019, Shah2026, JoHantgen}. The nature of any such objects and whether they are bound to the stars they occult remains a topic of ongoing investigation. Microlensing may offer a powerful technique to observe any such objects. If a J1407b-type object were acting as a gravitational lens, its rings would produce extensive complex structure in the lightcurve. The uncertainty surrounding the nature of the J1407b central object inspires us to briefly discuss another, much more hypothetical class of objects which could, in principle, host stable disks: primordial black holes.

\subsection{Primordial black holes}\label{holes}
Microlensing-derived abundance constraints on primordial black holes (PBHs) currently span planetary through stellar mass ranges \citep{CarrKuhnel2025}. If some PBHs retain rings or disks, the mechanism considered here could apply. A possible origin for circum-PBH disks is accretion after recombination and during the dark ages, a process that receives substantial attention because the resulting radiation would alter the observed cosmic microwave background (CMB) temperature anisotropy and polarization \citep{Ricottietal2008, AliHaimoud2017, Poulinetal2017, Serpico2020, DeLucaetal2020, Pigaetal2022, Facchinettietal2023, Agius, Jangra}. Although some calculations approximate the accretion flow as spherical, an idea that has gained consideration in the past decade is that captured gas carrying sufficient angular momentum would form a disk-like accretion flow, which would potentially lead to more stringent CMB bounds, although these constraints are highly model-dependent \citep{Poulinetal2017, Pigaetal2022, Facchinettietal2023, Agius, Jangra}. Existing studies primarily address how active disk-shaped accretion flows would impact CMB bounds on PBH abundance, as opposed to the question of whether the flow would completely accrete versus whether a cold, quiescent disk might ultimately persist over long timescales. Following initial accretion of baryonic matter, an isolated and effectively neutral black hole of the lens mass ranges targeted by surveys would itself supply negligible or zero intrinsic luminosity \citep{Hawking} or magnetic field \citep{Carter1971}; moreover, magnetohydrodynamic accretion mechanisms require the disk to remain sufficiently ionized \citep{BalbusHawley2000}, which might not occur over sufficiently long timescales to fully accrete disks in this setting. In the absence of specific mechanisms to produce effective viscosity, disks can, in principle, persist over long lifetimes \citep{Pringle1981}. It is intrinsically difficult to exclude or prove the long-term existence of such disks based on theory alone, leaving microlensing observations as a potentially important observational tool.

If such disks exist, they could consist largely of primordial baryonic matter, possibly including H$_2$-rich condensed solids. Whether solid para-H$_2$ ice can form and persist in astrophysical environments has been debated, but it is widely considered that such ice might exist in relatively cold and isolated settings \citep{Walker, Fuglistaler}. Characterization of the optical properties of solid para-H$_2$ previously suggested that extinction would be mostly due to scattering, with weak IR absorption features \citep{Kettwichetal2015}. In addition to extinction by scattering, hydrogen ice disks could affect microlensing lightcurves by refraction; laboratory measurements suggest solid para-H$_2$ has a refractive index of $\sim 1.13$ across the wavelength range $430-16700$ nm \citep{Pereraetal2011, Kettwichetal2015}, spanning the primary high-cadence microlensing band (F146) used by Roman's GBTDS. Because the angles of gravitational deflection typical in microlensing are minuscule, with the signal depending on precise deflections to reach the observer, a refractive index contrast of $\sim 0.13$ would be more than adequate to profoundly affect microlensing lightcurves as observed from Earth. Even if the hydrogen remained gaseous rather than condensing, it remains plausible that $H_2$-rich gas could adequately refract the lensed ray paths, since molecular hydrogen has measurable refractivity in the visible and near infrared \citep{PeckHuang1977}. However, whether such disks are expected to survive over galactic timescales and, if so, what their specific properties would be, remain unknown. Given the hypothesized importance of PBHs as a dark matter candidate, upcoming surveys such as Roman's GBTDS offer an exciting means to probe possible disks in these settings. 

\section{Conclusion} \label{sec:conc}
If upcoming surveys such as the Nancy Grace Roman Space Telescope Galactic Bulge Time-Domain Survey enable identification of just a single disk using this effect, it would offer an exciting application of microlensing and a novel probe of astrophysical disk dynamics. Moreover, it is crucial to consider the potential existence of such disks in lightcurve classification. As shown in this work, lightcurves produced from this effect can depart strongly from a Paczy\'{n}ski curve and could easily interfere with event classification if not accounted for in survey pipeline design. The lightcurve morphologies presented here provide a starting point for the design of such pipelines, which can be extended to account for more complex disks that include annular structures or partial attenuation.

The physics governing disk and ring dissipation depends on a complex interplay of dynamical processes, making the properties and long-term survival of these structures difficult to constrain in observation-limited settings. Accordingly, the prevalence of free-floating, disk-bearing objects remains highly uncertain. The abundances and properties of disks around free-floating planetary-mass objects, brown dwarfs, and putative primordial black holes hold broad relevance spanning cosmological and star/planet formation theory. Microlensing may be one of the few ways to constrain the abundance of such populations once their disks have cooled, making it especially important to account for the potential effects of disks in the design of microlensing surveys.

\begin{acknowledgments}
The author would like to thank R. Murray-Clay, J. Fortney, M. \'{C}uk, and W. DeRocco for encouraging discussions. Support during the early phases of this work was provided by the Heising-Simons Foundation 51 Pegasi b Fellowship. P. McGill provided substantial and helpful feedback, including pointing out the similarity of the Morphology [3,0] case to planetary caustics and the expected utility of paired astrometry for confirming and characterizing disk occultation events. OpenAI ChatGPT (GPT-5.5) and Google Gemini were used for limited assistance with editing pre-existing manuscript text and MATLAB code, and did not generate novel scientific insight. All resulting statements and calculations were validated by the author.
\end{acknowledgments}

%





\bibliography{sample631}{}

@ARTICLE{Agius,
       author = {{Agius}, Dominic and {Essig}, Rouven and {Gaggero}, Daniele and {Scarcella}, Francesca and {Suczewski}, Gregory and {Valli}, Mauro},
        title = "{Feedback in the dark: a critical examination of CMB bounds on primordial black holes}",
      journal = {\jcap},
         year = 2024,
        month = jul,
       volume = {2024},
       number = {7},
          eid = {003},
        pages = {003},
          doi = {10.1088/1475-7516/2024/07/003},
archivePrefix = {arXiv},
       eprint = {2403.18895},
 primaryClass = {hep-ph},
       adsurl = {https://ui.adsabs.harvard.edu/abs/2024JCAP...07..003A}
}

@ARTICLE{Alcocketal2000,
       author = {{Alcock}, C. and {Allsman}, R.~A. and {Alves}, D.~R. and {Axelrod}, T.~S. and {Becker}, A.~C. and {Bennett}, D.~P. and {Cook}, K.~H. and {Dalal}, N. and {Drake}, A.~J. and {Freeman}, K.~C. and {Geha}, M. and {Griest}, K. and {Lehner}, M.~J. and {Marshall}, S.~L. and {Minniti}, D. and {Nelson}, C.~A. and {Peterson}, B.~A. and {Popowski}, P. and {Pratt}, M.~R. and {Quinn}, P.~J. and {Stubbs}, C.~W. and {Sutherland}, W. and {Tomaney}, A.~B. and {Vandehei}, T. and {Welch}, D.},
        title = "{The MACHO Project: Microlensing Results from 5.7 Years of Large Magellanic Cloud Observations}",
      journal = {\apj},
         year = 2000,
        month = oct,
       volume = {542},
       number = {1},
        pages = {281-307},
          doi = {10.1086/309512},
archivePrefix = {arXiv},
       eprint = {astro-ph/0001272},
 primaryClass = {astro-ph},
       adsurl = {https://ui.adsabs.harvard.edu/abs/2000ApJ...542..281A}

}

@ARTICLE{AliHaimoud2017,
       author = {{Ali-Ha{\"\i}moud}, Yacine and {Kamionkowski}, Marc},
        title = "{Cosmic microwave background limits on accreting primordial black holes}",
      journal = {\prd},
         year = 2017,
        month = feb,
       volume = {95},
       number = {4},
          eid = {043534},
        pages = {043534},
          doi = {10.1103/PhysRevD.95.043534},
archivePrefix = {arXiv},
       eprint = {1612.05644},
 primaryClass = {astro-ph.CO},
       adsurl = {https://ui.adsabs.harvard.edu/abs/2017PhRvD..95d3534A}
}

@ARTICLE{Algol2002,
       author = {{Agol}, Eric},
        title = "{Occultation and Microlensing}",
      journal = {\apj},
         year = 2002,
        month = nov,
       volume = {579},
       number = {1},
        pages = {430-436},
          doi = {10.1086/342880},
archivePrefix = {arXiv},
       eprint = {astro-ph/0207228},
 primaryClass = {astro-ph},
       adsurl = {https://ui.adsabs.harvard.edu/abs/2002ApJ...579..430A}
}

@ARTICLE{BalbusHawley2000,
       author = {{Balbus}, Steven A. and {Hawley}, John F.},
        title = "{Solar Nebula Magnetohydrodynamics}",
      journal = {\ssr},
         year = 2000,
        month = apr,
       volume = {92},
        pages = {39-54},
          doi = {10.1023/A:1005293132737},
archivePrefix = {arXiv},
       eprint = {astro-ph/9906317},
 primaryClass = {astro-ph},
       adsurl = {https://ui.adsabs.harvard.edu/abs/2000SSRv...92...39B}
}

@ARTICLE{Bond2001,
       author = {{Bond}, I.~A. and {Abe}, F. and {Dodd}, R.~J. and {Hearnshaw}, J.~B. and {Honda}, M. and {Jugaku}, J. and {Kilmartin}, P.~M. and {Marles}, A. and {Masuda}, K. and {Matsubara}, Y. and {Muraki}, Y. and {Nakamura}, T. and {Nankivell}, G. and {Noda}, S. and {Noguchi}, C. and {Ohnishi}, K. and {Rattenbury}, N.~J. and {Reid}, M. and {Saito}, To. and {Sato}, H. and {Sekiguchi}, M. and {Skuljan}, J. and {Sullivan}, D.~J. and {Sumi}, T. and {Takeuti}, M. and {Watase}, Y. and {Wilkinson}, S. and {Yamada}, R. and {Yanagisawa}, T. and {Yock}, P.~C.~M.},
        title = "{Real-time difference imaging analysis of MOA Galactic bulge observations during 2000}",
      journal = {\mnras},
         year = 2001,
        month = nov,
       volume = {327},
       number = {3},
        pages = {868-880},
          doi = {10.1046/j.1365-8711.2001.04776.x},
archivePrefix = {arXiv},
       eprint = {astro-ph/0102181},
 primaryClass = {astro-ph},
       adsurl = {https://ui.adsabs.harvard.edu/abs/2001MNRAS.327..868B}
}

@ARTICLE{Bozzaetal2002,
       author = {{Bozza}, V. and {Jetzer}, Ph. and {Mancini}, L. and {Scarpetta}, G.},
        title = "{Microlensing by compact objects associated with gas clouds}",
      journal = {\aap},
         year = 2002,
        month = jan,
       volume = {382},
        pages = {6-16},
          doi = {10.1051/0004-6361:20011602},
archivePrefix = {arXiv},
       eprint = {astro-ph/0111079},
 primaryClass = {astro-ph},
       adsurl = {https://ui.adsabs.harvard.edu/abs/2002A&A...382....6B}
}

@ARTICLE{BozzaMancini2002,
       author = {{Bozza}, V. and {Mancini}, L.},
        title = "{Microlensing of strongly interacting binary systems}",
      journal = {\aap},
         year = 2002,
        month = nov,
       volume = {394},
        pages = {L47-L50},
          doi = {10.1051/0004-6361:20021409},
archivePrefix = {arXiv},
       eprint = {astro-ph/0209501},
 primaryClass = {astro-ph},
       adsurl = {https://ui.adsabs.harvard.edu/abs/2002A&A...394L..47B}
}

@ARTICLE{Bromley1996,
       author = {{Bromley}, B.~C.},
        title = "{Finite-Size Gravitational Microlenses}",
      journal = {\apj},
         year = 1996,
        month = aug,
       volume = {467},
        pages = {537},
          doi = {10.1086/177630},
       adsurl = {https://ui.adsabs.harvard.edu/abs/1996ApJ...467..537B}
}

@ARTICLE{CarrKuhnel2025,
       author = {{Carr}, Bernard and {Kuhnel}, Florian},
        title = "{Primordial Black Holes}",
      journal = {arXiv e-prints},
         year = 2025,
        month = feb,
          eid = {arXiv:2502.15279},
        pages = {arXiv:2502.15279},
          doi = {10.48550/arXiv.2502.15279},
archivePrefix = {arXiv},
       eprint = {2502.15279},
 primaryClass = {astro-ph.CO},
       adsurl = {https://ui.adsabs.harvard.edu/abs/2025arXiv250215279C}
}

@ARTICLE{Carter1971,
       author = {{Carter}, B.},
        title = "{Axisymmetric Black Hole Has Only Two Degrees of Freedom}",
      journal = {\prl},
         year = 1971,
        month = feb,
       volume = {26},
       number = {6},
        pages = {331-333},
          doi = {10.1103/PhysRevLett.26.331},
       adsurl = {https://ui.adsabs.harvard.edu/abs/1971PhRvL..26..331C}
}

@ARTICLE{Damianetal2025,
       author = {{Damian}, Belinda and {Scholz}, Aleks and {Jayawardhana}, Ray and {Almendros-Abad}, V. and {Flagg}, Laura and {Mu{\v{z}}i{\'c}}, Koraljka and {Natta}, Antonella and {Pinilla}, Paola and {Testi}, Leonardo},
        title = "{Spectroscopy of Free-floating Planetary-mass Objects and Their Disks with JWST}",
      journal = {\aj},
         year = 2025,
        month = aug,
       volume = {170},
       number = {2},
          eid = {127},
        pages = {127},
          doi = {10.3847/1538-3881/adea50},
archivePrefix = {arXiv},
       eprint = {2507.05155},
 primaryClass = {astro-ph.EP},
       adsurl = {https://ui.adsabs.harvard.edu/abs/2025AJ....170..127D}
}

@ARTICLE{DeLucaetal2020,
       author = {{De Luca}, V. and {Franciolini}, G. and {Pani}, P. and {Riotto}, A.},
        title = "{Constraints on primordial black holes: The importance of accretion}",
      journal = {\prd},
         year = 2020,
        month = aug,
       volume = {102},
       number = {4},
          eid = {043505},
        pages = {043505},
          doi = {10.1103/PhysRevD.102.043505},
archivePrefix = {arXiv},
       eprint = {2003.12589},
 primaryClass = {astro-ph.CO},
       adsurl = {https://ui.adsabs.harvard.edu/abs/2020PhRvD.102d3505D}
}

@ARTICLE{DeRocco2024,
       author = {{DeRocco}, William and {Frangipane}, Evan and {Hamer}, Nick and {Profumo}, Stefano and {Smyth}, Nolan},
        title = "{Revealing terrestrial-mass primordial black holes with the Nancy Grace Roman Space Telescope}",
      journal = {\prd},
         year = 2024,
        month = jan,
       volume = {109},
       number = {2},
          eid = {023013},
        pages = {023013},
          doi = {10.1103/PhysRevD.109.023013},
archivePrefix = {arXiv},
       eprint = {2311.00751},
 primaryClass = {astro-ph.CO},
       adsurl = {https://ui.adsabs.harvard.edu/abs/2024PhRvD.109b3013D}
}

@ARTICLE{Einstein1936,
       author = {{Einstein}, Albert},
        title = "{Lens-Like Action of a Star by the Deviation of Light in the Gravitational Field}",
      journal = {Science},
         year = 1936,
        month = dec,
       volume = {84},
       number = {2188},
        pages = {506-507},
          doi = {10.1126/science.84.2188.506},
       adsurl = {https://ui.adsabs.harvard.edu/abs/1936Sci....84..506E}
}

@ARTICLE{Facchinettietal2023,
       author = {{Facchinetti}, Ga{\'e}tan and {Lucca}, Matteo and {Clesse}, S{\'e}bastien},
        title = "{Relaxing CMB bounds on primordial black holes: The role of ionization fronts}",
      journal = {\prd},
         year = 2023,
        month = feb,
       volume = {107},
       number = {4},
          eid = {043537},
        pages = {043537},
          doi = {10.1103/PhysRevD.107.043537},
archivePrefix = {arXiv},
       eprint = {2212.07969},
 primaryClass = {astro-ph.CO},
       adsurl = {https://ui.adsabs.harvard.edu/abs/2023PhRvD.107d3537F}
}

@ARTICLE{Fores-Toribio,
       author = {{For{\'e}s-Toribio}, Raquel and {JoHantgen}, B. and {Kochanek}, C.~S. and {Jorstad}, S.~G. and {Hermes}, J.~J. and {Armstrong}, J.~D. and {Ashall}, C. and {Burns}, C.~R. and {Gaidos}, E. and {Hoogendam}, W.~B. and {Hsiao}, E.~Y. and {Medler}, K. and {Morrell}, N. and {Pfeffer}, C. and {Shappee}, B.~J. and {Stanek}, K. and {Tucker}, M.~A. and {Xiao}, H. and {Auchettl}, K. and {Lu}, L. and {Rowan}, D.~M. and {Vaccaro}, T. and {Williams}, J.~P.},
        title = "{ASASSN-24fw: An 8-month long, 4.1 mag, optically achromatic and polarized dimming event}",
      journal = {The Open Journal of Astrophysics},
         year = 2025,
        month = aug,
       volume = {8},
          eid = {114},
        pages = {114},
          doi = {10.33232/001c.143105},
archivePrefix = {arXiv},
       eprint = {2507.03080},
 primaryClass = {astro-ph.SR},
       adsurl = {https://ui.adsabs.harvard.edu/abs/2025OJAp....8E.114F}
}

@ARTICLE{Fuglistaler,
       author = {{F{\"u}glistaler}, A. and {Pfenniger}, D.},
        title = "{Solid H$_{2}$ in the interstellar medium}",
      journal = {\aap},
         year = 2018,
        month = jun,
       volume = {613},
          eid = {A64},
        pages = {A64},
          doi = {10.1051/0004-6361/201731739},
archivePrefix = {arXiv},
       eprint = {1712.01160},
 primaryClass = {astro-ph.GA},
       adsurl = {https://ui.adsabs.harvard.edu/abs/2018A&A...613A..64F}
}

@ARTICLE{Gaudietal2003,
       author = {{Gaudi}, B. Scott and {Chang}, Heon-Young and {Han}, Cheongho},
        title = "{Probing Structures of Distant Extrasolar Planets with Microlensing}",
      journal = {\apj},
         year = 2003,
        month = mar,
       volume = {586},
       number = {1},
        pages = {527-539},
          doi = {10.1086/367539},
archivePrefix = {arXiv},
       eprint = {astro-ph/0209512},
 primaryClass = {astro-ph},
       adsurl = {https://ui.adsabs.harvard.edu/abs/2003ApJ...586..527G}
}

@ARTICLE{Gould1992,
       author = {{Gould}, Andrew},
        title = "{Extending the MACHO Search to approximately 10 6 M sub sun}",
      journal = {\apj},
         year = 1992,
        month = jun,
       volume = {392},
        pages = {442},
          doi = {10.1086/171443},
       adsurl = {https://ui.adsabs.harvard.edu/abs/1992ApJ...392..442G}
}

@ARTICLE{GouldEscude1997,
       author = {{Gould}, Andrew and {Miralda-Escud{\'e}}, Jordi},
        title = "{Signatures of Accretion Disks in Quasar Microlensing}",
      journal = {\apjl},
         year = 1997,
        month = jul,
       volume = {483},
       number = {1},
        pages = {L13-L16},
          doi = {10.1086/310739},
archivePrefix = {arXiv},
       eprint = {astro-ph/9612144},
 primaryClass = {astro-ph},
       adsurl = {https://ui.adsabs.harvard.edu/abs/1997ApJ...483L..13G}
}

@ARTICLE{GouldLoeb,
       author = {{Gould}, Andrew and {Loeb}, Abraham},
        title = "{Discovering Planetary Systems through Gravitational Microlenses}",
      journal = {\apj},
         year = 1992,
        month = sep,
       volume = {396},
        pages = {104},
          doi = {10.1086/171700},
       adsurl = {https://ui.adsabs.harvard.edu/abs/1992ApJ...396..104G}
}

@ARTICLE{GraffGaudi2000,
       author = {{Graff}, David S. and {Gaudi}, B. Scott},
        title = "{Direct Detection of Large Close-in Planets around the Source Stars of Caustic-crossing Microlensing Events}",
      journal = {\apjl},
         year = 2000,
        month = aug,
       volume = {538},
       number = {2},
        pages = {L133-L136},
          doi = {10.1086/312811},
archivePrefix = {arXiv},
       eprint = {astro-ph/0004089},
 primaryClass = {astro-ph},
       adsurl = {https://ui.adsabs.harvard.edu/abs/2000ApJ...538L.133G}
}

@ARTICLE{GriestSafizadeh1998,
       author = {{Griest}, Kim and {Safizadeh}, Neda},
        title = "{The Use of High-Magnification Microlensing Events in Discovering Extrasolar Planets}",
      journal = {\apj},
         year = 1998,
        month = jun,
       volume = {500},
       number = {1},
        pages = {37-50},
          doi = {10.1086/305729},
archivePrefix = {arXiv},
       eprint = {astro-ph/9710342},
 primaryClass = {astro-ph},
       adsurl = {https://ui.adsabs.harvard.edu/abs/1998ApJ...500...37G}
}

@ARTICLE{HagenFian2026,
       author = {{Hagen}, Scott and {Fian}, Carina},
        title = "{Physically motivated AGN emissivity profiles and their effects on quasar microlensing signatures. 1. Multi-epoch accretion disc size inference}",
      journal = {arXiv e-prints},
         year = 2026,
        month = jul,
          eid = {arXiv:2607.03291},
        pages = {arXiv:2607.03291},
archivePrefix = {arXiv},
       eprint = {2607.03291},
 primaryClass = {astro-ph.GA},
       adsurl = {https://ui.adsabs.harvard.edu/abs/2026arXiv260703291H}
}

@ARTICLE{Hanetal2001,
       author = {{Han}, Cheongho and {Chang}, Heon-Young and {An}, Jin H. and {Chang}, Kyongae},
        title = "{Properties of microlensing light curve anomalies induced by multiple planets}",
      journal = {\mnras},
         year = 2001,
        month = dec,
       volume = {328},
       number = {3},
        pages = {986-992},
          doi = {10.1046/j.1365-8711.2001.04973.x},
archivePrefix = {arXiv},
       eprint = {astro-ph/0107517},
 primaryClass = {astro-ph},
       adsurl = {https://ui.adsabs.harvard.edu/abs/2001MNRAS.328..986H}
}

@ARTICLE{Hawking,
       author = {{Hawking}, S.~W.},
        title = "{Particle creation by black holes}",
      journal = {Communications in Mathematical Physics},
         year = 1975,
        month = aug,
       volume = {43},
       number = {3},
        pages = {199-220},
          doi = {10.1007/BF02345020},
       adsurl = {https://ui.adsabs.harvard.edu/abs/1975CMaPh..43..199H}
}

@ARTICLE{Hundertmark,
       author = {{Hundertmark}, M. and {Hessman}, F.~V. and {Dreizler}, S.},
        title = "{Detecting circumstellar disks around gravitational microlenses}",
      journal = {\aap},
         year = 2009,
        month = jun,
       volume = {500},
       number = {2},
        pages = {929-934},
          doi = {10.1051/0004-6361/200811458},
archivePrefix = {arXiv},
       eprint = {0904.1117},
 primaryClass = {astro-ph.EP},
       adsurl = {https://ui.adsabs.harvard.edu/abs/2009A&A...500..929H}
}

@ARTICLE{Jangra,
       author = {{Jangra}, Pratibha and {Gaggero}, Daniele and {Kavanagh}, Bradley J. and {Diego}, J.~M.},
        title = "{The cosmic history of Primordial Black Hole accretion and its uncertainties}",
      journal = {\jcap},
         year = 2025,
        month = aug,
       volume = {2025},
       number = {8},
          eid = {006},
        pages = {006},
          doi = {10.1088/1475-7516/2025/08/006},
archivePrefix = {arXiv},
       eprint = {2412.11921},
 primaryClass = {astro-ph.CO},
       adsurl = {https://ui.adsabs.harvard.edu/abs/2025JCAP...08..006J}
}

@ARTICLE{JoHantgen,
       author = {{JoHantgen}, B. and {Rowan}, D.~M. and {For{\'e}s-Toribio}, R. and {Kochanek}, C.~S. and {Stanek}, K.~Z. and {Shappee}, B.~J. and {Dong}, Subo and {Prieto}, J.~L. and {Thompson}, Todd A.},
        title = "{A Systematic Search for Big Dippers in ASAS-SN}",
      journal = {The Open Journal of Astrophysics},
         year = 2026,
        month = feb,
       volume = {9},
        pages = {56224},
          doi = {10.33232/001c.156224},
archivePrefix = {arXiv},
       eprint = {2507.19594},
 primaryClass = {astro-ph.SR},
       adsurl = {https://ui.adsabs.harvard.edu/abs/2026OJAp....956224J}
}

@ARTICLE{Johnsonetal2020,
       author = {{Johnson}, Samson A. and {Penny}, Matthew and {Gaudi}, B. Scott and {Kerins}, Eamonn and {Rattenbury}, Nicholas J. and {Robin}, Annie C. and {Calchi Novati}, Sebastiano and {Henderson}, Calen B.},
        title = "{Predictions of the Nancy Grace Roman Space Telescope Galactic Exoplanet Survey. II. Free-floating Planet Detection Rates}",
      journal = {\aj},
         year = 2020,
        month = sep,
       volume = {160},
       number = {3},
          eid = {123},
        pages = {123},
          doi = {10.3847/1538-3881/aba75b},
archivePrefix = {arXiv},
       eprint = {2006.10760},
 primaryClass = {astro-ph.EP},
       adsurl = {https://ui.adsabs.harvard.edu/abs/2020AJ....160..123J}
}

@ARTICLE{KenworthyMamajek2015,
       author = {{Kenworthy}, M.~A. and {Mamajek}, E.~E.},
        title = "{Modeling Giant Extrasolar Ring Systems in Eclipse and the Case of J1407b: Sculpting by Exomoons?}",
      journal = {\apj},
         year = 2015,
        month = feb,
       volume = {800},
       number = {2},
          eid = {126},
        pages = {126},
          doi = {10.1088/0004-637X/800/2/126},
archivePrefix = {arXiv},
       eprint = {1501.05652},
 primaryClass = {astro-ph.SR},
       adsurl = {https://ui.adsabs.harvard.edu/abs/2015ApJ...800..126K}
}

@ARTICLE{Kenworthyetal2015,
       author = {{Kenworthy}, M.~A. and {Lacour}, S. and {Kraus}, A. and {Triaud}, A.~H.~M.~J. and {Mamajek}, E.~E. and {Scott}, E.~L. and {S{\'e}gransan}, D. and {Ireland}, M. and {Hambsch}, F.-J. and {Reichart}, D.~E. and {Haislip}, J.~B. and {LaCluyze}, A.~P. and {Moore}, J.~P. and {Frank}, N.~R.},
        title = "{Mass and period limits on the ringed companion transiting the young star J1407}",
      journal = {\mnras},
         year = 2015,
        month = jan,
       volume = {446},
       number = {1},
        pages = {411-427},
          doi = {10.1093/mnras/stu2067},
archivePrefix = {arXiv},
       eprint = {1410.6577},
 primaryClass = {astro-ph.SR},
       adsurl = {https://ui.adsabs.harvard.edu/abs/2015MNRAS.446..411K}
}

@ARTICLE{Kenworthyetal2020,
       author = {{Kenworthy}, M.~A. and {Klaassen}, P.~D. and {Min}, M. and {van der Marel}, N. and {Bohn}, A.~J. and {Kama}, M. and {Triaud}, A. and {Hales}, A. and {Monkiewicz}, J. and {Scott}, E. and {Mamajek}, E.~E.},
        title = "{ALMA and NACO observations towards the young exoring transit system J1407 (V1400 Cen)}",
      journal = {\aap},
         year = 2020,
        month = jan,
       volume = {633},
          eid = {A115},
        pages = {A115},
          doi = {10.1051/0004-6361/201936141},
archivePrefix = {arXiv},
       eprint = {1912.03314},
 primaryClass = {astro-ph.EP},
       adsurl = {https://ui.adsabs.harvard.edu/abs/2020A&A...633A.115K}
}

@ARTICLE{Kettwichetal2015,
       author = {{Kettwich}, Sharon C. and {Anderson}, David T. and {Walker}, Mark A. and {Tuntsov}, Artem V.},
        title = "{The infrared dielectric function of solid para-hydrogen}",
      journal = {\mnras},
         year = 2015,
        month = jun,
       volume = {450},
       number = {1},
        pages = {1032-1041},
          doi = {10.1093/mnras/stv691},
archivePrefix = {arXiv},
       eprint = {1503.05257},
 primaryClass = {astro-ph.GA},
       adsurl = {https://ui.adsabs.harvard.edu/abs/2015MNRAS.450.1032K}
}

@ARTICLE{Klaassen,
       author = {{Klaassen}, Pamela and {Kenworthy}, Matthew A. and {Mamajek}, Eric E. and {van der Marel}, Nienke and {Min}, Michiel and {Triaud}, Amaury H.~M.~J. and {Hales}, Antonio S.},
        title = "{Non-detection of J1407 b in ALMA Band 7 Observations}",
      journal = {Research Notes of the American Astronomical Society},
         year = 2025,
        month = dec,
       volume = {9},
       number = {12},
          eid = {326},
        pages = {326},
          doi = {10.3847/2515-5172/ae2680},
archivePrefix = {arXiv},
       eprint = {2512.13254},
 primaryClass = {astro-ph.GA},
       adsurl = {https://ui.adsabs.harvard.edu/abs/2025RNAAS...9..326K}
}

@ARTICLE{Kochanek2004,
       author = {{Kochanek}, C.~S.},
        title = "{Quantitative Interpretation of Quasar Microlensing Light Curves}",
      journal = {\apj},
         year = 2004,
        month = apr,
       volume = {605},
       number = {1},
        pages = {58-77},
          doi = {10.1086/382180},
archivePrefix = {arXiv},
       eprint = {astro-ph/0307422},
 primaryClass = {astro-ph},
       adsurl = {https://ui.adsabs.harvard.edu/abs/2004ApJ...605...58K}
}

@ARTICLE{Langeveld,
       author = {{Langeveld}, Adam B. and {Scholz}, Aleks and {Mu{\v{z}}i{\'c}}, Koraljka and {Jayawardhana}, Ray and {Capela}, Daniel and {Albert}, Lo{\"\i}c and {Doyon}, Ren{\'e} and {Flagg}, Laura and {de Furio}, Matthew and {Johnstone}, Doug and {Lafr{\`e}niere}, David and {Meyer}, Michael},
        title = "{The JWST/NIRISS Deep Spectroscopic Survey for Young Brown Dwarfs and Free-floating Planets}",
      journal = {\aj},
         year = 2024,
        month = oct,
       volume = {168},
       number = {4},
          eid = {179},
        pages = {179},
          doi = {10.3847/1538-3881/ad6f0c},
archivePrefix = {arXiv},
       eprint = {2408.12639},
 primaryClass = {astro-ph.EP},
       adsurl = {https://ui.adsabs.harvard.edu/abs/2024AJ....168..179L}
}

@ARTICLE{LewisIbata2000,
       author = {{Lewis}, Geraint F. and {Ibata}, Rodrigo A.},
        title = "{Probing the Atmospheres of Planets Orbiting Microlensed Stars via Polarization Variability}",
      journal = {\apjl},
         year = 2000,
        month = aug,
       volume = {539},
       number = {1},
        pages = {L63-L66},
          doi = {10.1086/312826},
archivePrefix = {arXiv},
       eprint = {astro-ph/0006261},
 primaryClass = {astro-ph},
       adsurl = {https://ui.adsabs.harvard.edu/abs/2000ApJ...539L..63L}
}

@ARTICLE{Mamajeketal2012,
       author = {{Mamajek}, Eric E. and {Quillen}, Alice C. and {Pecaut}, Mark J. and {Moolekamp}, Fred and {Scott}, Erin L. and {Kenworthy}, Matthew A. and {Collier Cameron}, Andrew and {Parley}, Neil R.},
        title = "{Planetary Construction Zones in Occultation: Discovery of an Extrasolar Ring System Transiting a Young Sun-like Star and Future Prospects for Detecting Eclipses by Circumsecondary and Circumplanetary Disks}",
      journal = {\aj},
         year = 2012,
        month = mar,
       volume = {143},
       number = {3},
          eid = {72},
        pages = {72},
          doi = {10.1088/0004-6256/143/3/72},
archivePrefix = {arXiv},
       eprint = {1108.4070},
 primaryClass = {astro-ph.SR},
       adsurl = {https://ui.adsabs.harvard.edu/abs/2012AJ....143...72M}
}

@ARTICLE{MaoPaczynski1991,
       author = {{Mao}, Shude and {Paczy\'{n}ski}, Bohdan},
        title = "{Gravitational Microlensing by Double Stars and Planetary Systems}",
      journal = {\apjl},
         year = 1991,
        month = jun,
       volume = {374},
        pages = {L37},
          doi = {10.1086/186066},
       adsurl = {https://ui.adsabs.harvard.edu/abs/1991ApJ...374L..37M}
}

@ARTICLE{Menteletal2018,
       author = {{Mentel}, R.~T. and {Kenworthy}, M.~A. and {Cameron}, D.~A. and {Scott}, E.~L. and {Mellon}, S.~N. and {Hudec}, R. and {Birkby}, J.~L. and {Mamajek}, E.~E. and {Schrimpf}, A. and {Reichart}, D.~E. and {Haislip}, J.~B. and {Kouprianov}, V.~V. and {Hambsch}, F.-J. and {Tan}, T.-G. and {Hills}, K. and {Grindlay}, J.~E. and {Rodriguez}, J.~E. and {Lund}, M.~B. and {Kuhn}, R.~B.},
        title = "{Constraining the period of the ringed secondary companion to the young star J1407 with photographic plates}",
      journal = {\aap},
         year = 2018,
        month = nov,
       volume = {619},
          eid = {A157},
        pages = {A157},
          doi = {10.1051/0004-6361/201834004},
archivePrefix = {arXiv},
       eprint = {1810.05171},
 primaryClass = {astro-ph.EP},
       adsurl = {https://ui.adsabs.harvard.edu/abs/2018A&A...619A.157M}
}

@ARTICLE{Nieuwenhuizen,
       author = {{Nieuwenhuizen}, Theodorus Maria},
        title = "{A partially occulting MACHO-microlensing event in the Twin Quasar Q0957+561}",
      journal = {Fortschritte der Physik},
         year = 2017,
        month = jun,
       volume = {65},
       number = {6-8},
        pages = {1600107},
          doi = {10.1002/prop.201600107},
archivePrefix = {arXiv},
       eprint = {1612.08919},
 primaryClass = {astro-ph.GA},
       adsurl = {https://ui.adsabs.harvard.edu/abs/2017ForPh..6500107N}
}

@ARTICLE{Paczynski,
       author = {{Paczy\'{n}ski}, B.},
        title = "{Gravitational Microlensing by the Galactic Halo}",
      journal = {\apj},
         year = 1986,
        month = may,
       volume = {304},
        pages = {1},
          doi = {10.1086/164140},
       adsurl = {https://ui.adsabs.harvard.edu/abs/1986ApJ...304....1P}
}

@ARTICLE{Paczynski1996,
       author = {{Paczy\'{n}ski}, Bohdan},
        title = "{Gravitational Microlensing in the Local Group}",
      journal = {\araa},
         year = 1996,
        month = jan,
       volume = {34},
        pages = {419-460},
          doi = {10.1146/annurev.astro.34.1.419},
archivePrefix = {arXiv},
       eprint = {astro-ph/9604011},
 primaryClass = {astro-ph},
       adsurl = {https://ui.adsabs.harvard.edu/abs/1996ARA&A..34..419P}
}

@ARTICLE{PeckHuang1977,
       author = {{Peck}, E.~R. and {Huang}, S.},
        title = "{Refractivity and dispersion of hydrogen in the visible and near infrared (E)}",
      journal = {Journal of the Optical Society of America (1917-1983)},
         year = 1977,
        month = nov,
       volume = {67},
        pages = {1550},
          doi = {10.1364/JOSA.67.001550},
       adsurl = {https://ui.adsabs.harvard.edu/abs/1977JOSA...67.1550P}
}
\bibliographystyle{aasjournal}



\end{document}